# Moment Closure Stability Analysis of Stochastic Reaction Networks with Oscillatory Dynamics

Pedro H. Constantino*, Yiannis N. Kaznessis†

a Department of Chemical Engineering and Materials Science, University of Minnesota, 421 Washington Ave. SE, Minneapolis, MN 55455, USA.

* Correspondence: const071@umn.edu

† Current address: General Probiotics Inc., St. Paul, MN 55114, USA; yiannis@umn.edu.

**Abstract**

Biochemical reactions with oscillatory behavior play an essential role in synthetic biology at the microscopic scale. Although a robust stability theory for deterministic chemical oscillators in the macroscopic limit exists, the dynamical stability of stochastic oscillators is an object of ongoing research. The Chemical Master Equation along with kinetic Monte Carlo simulations constitute the most accurate approach to modeling microscopic systems. However, because of the challenges of solving the fully probabilistic model, most studies in stability analysis have focused on the description of externally disturbed oscillators. Here we apply the Maximum Entropy Principle as closure criterion for moment equations of oscillatory networks and perform the stability analysis of the internally disturbed Brusselator network. Particularly, we discuss the effects of kinetic and size parameters on the dynamics of this stochastic oscillatory system with intrinsic noise. Our numerical experiments reveal that changes in kinetic parameters lead to phenomenological and dynamical Hopf bifurcations, while reduced system sizes in the oscillatory region can reverse the stochastic Hopf dynamical bifurcations at the ensemble level. This is a unique feature of the stochastic dynamics of oscillatory systems, with unknown parallel in the macroscopic limit.

1. **Introduction**

For over 60 years now, chemical oscillators have received increased attention as new experimental discoveries are made in different biological contexts such as metabolism, signaling and cell development [1]. These complex systems can be mathematically modeled as chemical reaction networks through mappings of biomolecular interactions that translate into sets of nonlinear differential equations. The main concern in modeling the dynamics of such systems is to determine the conditions that promote oscillatory behavior and verify stability of stationary solutions to parametric changes or perturbations in the system.

The Brusselator is a benchmark model that provides simple qualitative description of biochemical clocks through a hypothetical set of chemical reactions. The system transitions from a stable solution to sustained oscillations when kinetic parameters are suitably varied [2-5]. The stability analysis describes the deterministic behavior as a Hopf bifurcation, where the stability of the solution changes because the pair of eigenvalues of the Jacobian matrix of the dynamical system crosses the imaginary axis [4].

Chemical oscillators in biological systems, however, are faced with at least two categories of noise that confer a stochastic nature on the reaction network. First, there are large variations across cell populations because each cell is a unique individual that may be at a different life-stage of the cell cycle. This main feature along with fluctuations in the environment constitute the external sources of noise. Secondly, for most biological applications the number of molecules interacting in reaction networks inside each cell is significantly small (e.g., in transcription there may be a single copy of a DNA molecule). Bioluminescence in flow cytometry experiments, for instance, is the collective result of thousands of individual cells which are strongly affected by intrinsic cellular noise [6]. Therefore, the internal fluctuations or the inherent stochasticity of the reaction system must also be accounted.

The stability of the externally disturbed Brusselator has been extensively explored in the literature by Arnold et al. [7], who modeled the external noise through Stochastic Differential Equations. Using a Langevin-Itô approach, these authors demonstrated that a parametric additive white noise destroys Hopf bifurcations. The topological change in the probability distribution due to different parametric regions is what Arnold coined as a phenomenological bifurcation (P-bifurcation) [8]. Arnold and co-workers proved that the Brusselator system does not go through a proper dynamical bifurcation (D-bifurcation), meaning that the sign of Lyapunov exponents of probability measures remains unchanged [7-8].

More recently, Bashkirtseva et al. showed that multiplicative parametric noise (i.e., noise such that the undisturbed solutions also solve the disturbed problem) can also induce a delaying shift of the Hopf bifurcation point [9]. In this case, a phenomenon of reverse stochastic bifurcation was observed in which oscillations were suppressed by the multiplicative noise. The numerical-analytical investigation was conducted using both the canonical Hopf model and the van der Pol oscillator. The bifurcation shift was estimated as proportional to the square of the noise intensity.

Little effort, however, to the extent of our knowledge, has gone to develop a framework for the stability of stochastic chemical reaction networks specifically accounting for internal fluctuations. The solution of the Chemical Master Equation, which constitutes the most accurate approach for modeling the internal noise of biochemical reaction networks, remains the main roadblock to this development [10]. However, a recent use of the principle of maximum information entropy has been successfully applied as moment closure for stochastic networks with oscillatory behavior [17]. Hence, extending such approach, in the next section we introduce the Brusselator model, the Chemical Master Equation method, the information entropy approach and the stability analysis of the stochastic Brusselator. In section 3, we demonstrate that the stochastic Brusselator undergoes both a phenomenological and dynamical Hopf bifurcation when kinetic parameters are varied. We also observe a unique reversal of the stochastic Hopf effect by reducing the system size, which is unknown in the macroscopic limit. Finally, we conclude the paper summarizing our findings and remarking potential applications.

## 2. Theoretical Background

### 2.1 The Brusselator Model

The Brusselator model is described by the following set of elementary chemical reaction steps:

$$A \xrightarrow{k_1'} X \tag{1.a}$$

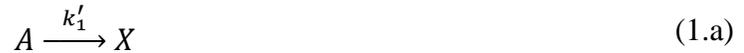

$$2X + Y \xrightarrow{k_2'} 3X \tag{1.b}$$

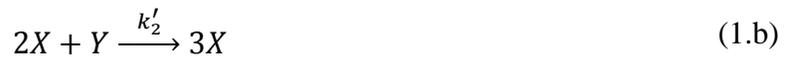

$$B + X \xrightarrow{k_3'} Y + C \tag{1.c}$$

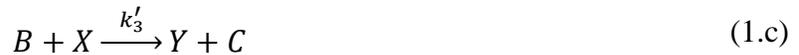

$$X \xrightarrow{k_4'} D \tag{1.d}$$

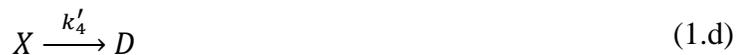

The concentrations of input reactants A and B are kept constant because they are assumed to be chemical reservoirs. Therefore, the model characterizes an open system far from its equilibrium state. If we define $k_1 = k_1'A$, $k_2 = k_2'$, $k_3 = k_3'B$, and $k_4 = k_4'$, one is able to apply the law of mass action to each elementary reaction step and derive a set of coupled ODEs for the concentrations of the intermediate species X and Y:

$$\frac{dX}{dt} = k_1 - (k_3 + k_4)X + k_2 X^2 Y \tag{2.a}$$

$$\frac{dY}{dt} = k_3 X - k_2 X^2 Y \tag{2.b}$$

The autocatalytic step of the intermediate X provides an important feature of this oscillatory system which is its feedback nature [1-3].

*2.2 Chemical Master Equation of the Brusselator*

The Chemical Master Equation for an isothermal chemical reaction network with $N$ species and $M$ reactions at constant volume is given by Eq. (9), which describes the probability distribution $\mathcal{P}(X, t)$ of the molecular populations $X(t)$ over time. Based on the concept of conservation of probability, the CME can be interpreted as a balance equation between arriving and leaving each state of the system [11-13]

$$\frac{\partial \mathcal{P}(X, t)}{\partial t} = \sum_{j=1}^{M} \left[ a_j(X - v_j) \mathcal{P}(X - v_j, t) - a_j(X) \mathcal{P}(X, t) \right] \tag{9}$$

The probability density that reaction $j$ will occur in the next infinitesimal time interval is expressed by the propensity function $a_j(X)$, while the stoichiometric matrix $v$ ($N \times M$) defines the state-change that results from each reaction occurring [10].

For the Brusselator reaction network the Chemical Master Equation is given by:

$$\begin{aligned}\frac{\partial \mathcal{P}(X,Y,t)}{\partial t} &= k_1 \mathcal{P}(X-1, Y, t) + k_2(X-1)(X-2)(Y+1)\mathcal{P}(X-1, Y+1, t) \\ &+ k_3(X+1)\mathcal{P}(X+1, Y-1, t) + k_4(X+1)\mathcal{P}(X+1, Y, t) \\ &- [k_1 + k_2 X(X-1)Y + k_3 X + k_4 X]\mathcal{P}(X, Y, t)\end{aligned} \tag{10}$$

The CME for the Brusselator is analytically unsolvable. Eq. (10) is an implicit function of the probability evaluated at different points of the phase space and, therefore, constitutes an infinite set of coupled equations.

*2.3 Moment Equation and Maximum Entropy Principle*

The CME can be alternatively written in terms of probability moments, which are the expectation values of its random variables [14]. This mathematical transformation produces a set of differential equations describing the dynamics of such moments ($\boldsymbol{\mu}$) in terms of higher order moments ($\boldsymbol{\mu}'$):

$$\frac{d\boldsymbol{\mu}}{dt} = \boldsymbol{\mu_0} + A\boldsymbol{\mu} + A'\boldsymbol{\mu}' \tag{11}$$

where $\boldsymbol{\mu_0}$ is a constant vector. Efficient computer algorithms have been developed to generate the coefficient matrices $A$ and $A'$ for arbitrary networks [15]. However, the dynamics of the lower order moments is underspecified and the system needs to be truncated in order to be solved. Our group created a method based on the maximization of the entropy of the probability distribution to calculate the higher order ones [16].

The maximum entropy distribution $P_H(X, Y)$ for a discrete bivariate system is given by Eq. (12), where $N_m$ is the number of lower-order moments, $\lambda_i$ is the Lagrange multiplier and $f_{\mu_i}$ is the functional form of the $i$-th lower moment (i.e., $\mu_i = E[f_{\mu_i}(\boldsymbol{X})]$, so that $\mu_1 = \langle X \rangle = E[X]$ and $f_{\mu_1} = X$, for instance) [17]:

$$P_H(\boldsymbol{X}, \boldsymbol{\lambda}) = \exp\left(-\lambda_0 - \sum_{k=1}^{N_m} \lambda_k f_{\mu_k}(\boldsymbol{X})\right) \tag{12}$$

Let $\mathfrak{D} = \{X \in \mathbb{Z}^{*N} \mid X \text{ is an available state}\}$, where $\mathbb{Z}^*$ represents the set of nonnegative integers. If we impose the following restriction to normalize the probability distribution,

$$\mu_0 = \sum_{\mathfrak{D}} P_H(X, \lambda) = 1 \tag{13}$$

Then we conclude that:

$$\lambda_0 = \ln\left\{\sum_{\mathfrak{D}} \exp\left(-\sum_{k=1}^{N_m} \lambda_k f_{\mu_k}(X)\right)\right\} \tag{14}$$

We then can write the final form of the probability function as:

$$P_H(X, \lambda) = \exp\left(-\sum_{k=1}^{N_m} \lambda_k f_{\mu_k}(X)\right)\left\{\sum_{\mathfrak{D}} \exp\left(-\sum_{k=1}^{N_m} \lambda_k f_{\mu_k}(X)\right)\right\}^{-1} \tag{15}$$

From Eq. (14), it is also evident that

$$\frac{\partial \lambda_0}{\partial \lambda_i} = \mu_i \tag{16}$$

From Eq. (15) one is then able to calculate the higher-order moments:

$$\mu'_{H_i} = \sum_{\mathfrak{D}} f_{\mu'_i}(X, Y) \mathcal{P}_H(X, Y) \tag{17}$$

By substituting Eq. (17) in Eq. (9), we arrive at the closed system of moment equations:

$$\frac{d\boldsymbol{\mu}_H(\lambda)}{dt} = \boldsymbol{\mu_0} + A\boldsymbol{\mu}_H(\lambda) + A'\boldsymbol{\mu}'_H(\lambda) \tag{18}$$

This approach has been validated for small networks of nonlinear chemical reactions with simple dynamics [16, 18] and for oscillatory systems [17].

At equilibrium, we have:

$$F(\lambda) = \mu_0 + A\mu_H(\lambda) + A'\mu'(\lambda) = 0 \tag{19}$$

A nonlinear system of $N_m$ algebraic equations that can be solved through Newton-Raphson iterations:

$$\lambda_{new} = \lambda_{old} - J_{NR}^{-1}F(\lambda) \tag{20}$$

where:

$$J_{NR} = \frac{\partial F(\lambda)}{\partial \lambda} = A\left(\frac{\partial \mu_H}{\partial \lambda}\right) + A'\left(\frac{\partial \mu'_H}{\partial \lambda}\right) \tag{21}$$

Here $\left(\frac{\partial \mu_H}{\partial \lambda}\right)$ is a $N_m \times N_m$ square matrix, whose elements are given by:

$$\frac{\partial \mu_{H_i}}{\partial \lambda_j} = \sum_{\mathfrak{D}} f_{\mu_i}(X) \left[ -\frac{f_{\mu_j} \exp(-\sum_{k=1}^{N_m} \lambda_k f_{\mu_k}(X))}{\sum_{\mathfrak{D}} \exp(-\sum_{k=1}^{N_m} \lambda_k f_{\mu_k}(X))} \right. \\ \left. + \exp\left(-\sum_{k=1}^{N_m} \lambda_k f_{\mu_k}(X)\right) \frac{\sum_{\mathfrak{D}} f_{\mu_j}(X) \exp(-\sum_{k=1}^{N_m} \lambda_k f_{\mu_k}(X))}{\{\sum_{\mathfrak{D}} \exp(-\sum_{k=1}^{N_m} \lambda_k f_{\mu_k}(X))\}^2} \right] \tag{22}$$

which is simplified to:

$$\frac{\partial \mu_{H_i}}{\partial \lambda_j} = \mu_i \mu_j - \sum_{\mathfrak{D}} \frac{f_{\mu_i}(X) f_{\mu_j}(X) \exp(-\sum_{k=1}^{N_m} \lambda_k f_{\mu_k}(X))}{\sum_{\mathfrak{D}} \exp(-\sum_{k=1}^{N_m} \lambda_k f_{\mu_k}(X))} \tag{23}$$

It is also important to observe from Eqs. (16) and (23) that:

$$\frac{\partial \mu_{H_i}}{\partial \lambda_j} = \frac{\partial^2 \lambda_0}{\partial \lambda_i \partial \lambda_j} = \frac{\partial \mu_{H_j}}{\partial \lambda_i} \tag{24}$$

Therefore, $\left(\frac{\partial \mu_H}{\partial \lambda}\right)$ is a symmetric matrix with real components (Hessian matrix of $\lambda_0$). Consequently, from fundamental theorems of linear algebra, we know that $\left(\frac{\partial \mu_H}{\partial \lambda}\right)$ can be diagonalized by a unitary matrix ($Q$) whose columns are its eigenvectors and the components of the resulting diagonal matrix ($\Lambda$) are its eigenvalues:

$$Q^T \left(\frac{\partial \boldsymbol{\mu_H}}{\partial \boldsymbol{\lambda}}\right) Q = \Lambda \Leftrightarrow \left(\frac{\partial \boldsymbol{\mu_H}}{\partial \boldsymbol{\lambda}}\right) = Q\Lambda Q^T \quad (25)$$

Hence,

$$\left(\frac{\partial \boldsymbol{\mu_H}}{\partial \boldsymbol{\lambda}}\right)^{-1} = (Q\Lambda Q^T)^{-1} = (Q^T)^{-1}\Lambda^{-1}Q^{-1} = Q\Lambda^{-1}Q^T \quad (26)$$

Similar to $\left(\frac{\partial \boldsymbol{\mu_H}}{\partial \boldsymbol{\lambda}}\right)$, the elements of $\left(\frac{\partial \boldsymbol{\mu'_H}}{\partial \boldsymbol{\lambda}}\right)$ are determined by:

$$\frac{\partial \mu'_{H_i}}{\partial \lambda_j} = \mu'_i \mu_j - \sum_{\mathfrak{D}} \frac{f_{\mu'_i}(\boldsymbol{X}) f_{\mu_j}(\boldsymbol{X}) \exp\left(-\sum_{k=1}^{N_m} \lambda_k f_{\mu_k}(\boldsymbol{X})\right)}{\sum_{\mathfrak{D}} \exp\left(-\sum_{k=1}^{N_m} \lambda_k f_{\mu_k}(\boldsymbol{X})\right)} \quad (27)$$

Consequently, we now can compute the Lagrange multipliers using Eq. (19) and the Maximum Entropy probability distribution using Eq. (11). These calculations will prove useful in the next section.

*2.4 Stability Analysis of the Stochastic Brusselator with Internal Noise*

The ability to cast the Chemical Master Equation into a deterministic system of moment equations and obtain its stationary solution facilitates the performance of stability analysis for intrinsically stochastic systems. Smadbeck and Kaznessis proposed the linearization of the dynamical system of probability moments near the equilibrium state through the computation of its Jacobian ($J_{eq}$) matrix [19]. Hence, by simply carrying out a Taylor expansion of the right side of Eq. (9) around the equilibrium state, we can use the following linear approximation:

$$\frac{d\boldsymbol{\mu}}{dt} \approx J_{eq}(\boldsymbol{\mu} - \boldsymbol{\mu_{eq}}) \quad (28)$$

Neglecting higher order terms, we can evaluate the local stability of the equilibrium state solution through the analysis of eigenvalues and eigenvectors of the Jacobian matrix, which can be obtained from:

$$J_{eq} = A + A' \frac{\partial \boldsymbol{\mu}'}{\partial \boldsymbol{\mu}}\bigg|_{eq} \tag{29}$$

Now, if the maximum entropy closure scheme is applied, both higher and lower-order moments are related to each other by the Lagrange multipliers of the distribution. In this case, the chain rule for partial derivatives and some differential manipulation can be used to expand the last term in Eq. (29):

$$\frac{\partial \boldsymbol{\mu}'_H}{\partial \boldsymbol{\mu}_H}\bigg|_{eq} = \left(\frac{\partial \boldsymbol{\mu}'_H}{\partial \boldsymbol{\lambda}}\right)\bigg|_{eq} \left(\frac{\partial \boldsymbol{\lambda}}{\partial \boldsymbol{\mu}_H}\right)\bigg|_{eq} = \left(\frac{\partial \boldsymbol{\mu}'_H}{\partial \boldsymbol{\lambda}}\bigg|_{eq}\right)\left(\frac{\partial \boldsymbol{\mu}_H}{\partial \boldsymbol{\lambda}}\bigg|_{eq}\right)^{-1} \tag{30}$$

The last equality comes from the Inverse Function Theorem applied to multivariable vector functions, where we assume that the Jacobian determinant of the lower moments is nonzero, so that the inverse matrix necessarily exists. Therefore, based on the entropy closure method, here we propose a simple final expression for the equilibrium state Jacobian matrix combining Eqs. (26), (29) and (30):

$$J_{eq} = A + A' \left(\frac{\partial \boldsymbol{\mu}'_H}{\partial \boldsymbol{\lambda}}\bigg|_{eq}\right)\left(\frac{\partial \boldsymbol{\mu}_H}{\partial \boldsymbol{\lambda}}\bigg|_{eq}\right)^{-1} = A + A' \left(\frac{\partial \boldsymbol{\mu}'_H}{\partial \boldsymbol{\lambda}}\bigg|_{eq}\right) Q \Lambda^{-1} Q^{\mathrm{T}} \tag{31}$$

whereby we may compute the eigenvalues and eigenvectors of the system of moment equations.

Next, we apply these calculations to the Brusselator, to provide support for a framework of stability analysis of stochastic oscillatory chemical reaction networks.

## 3. Results and Discussion

In this section we explore the kinetic and size effects on the stability and dynamics of the internally stochastic Brusselator. The kinetic effects are tested by simulating systems with different $k_2\ [molec^{-2} sec^{-1}]$ values while keeping constant $k_1 = 625\ molec.sec^{-1}, k_3 = 50\ sec^{-1}$, $k_4 = 5\ sec^{-1}$. The size effects are quantified by the size parameter $\Omega$. From a straightforward dimension analysis, if two Brusselator systems are such that $k_1^* = \Omega\ k_1$ and $k_2^* = \Omega^{-2}\ k_2$, then their volumes are such that $V^* = \Omega\ V$. Therefore, a sequence of Brusselator systems following this mathematical transformation will simulate the effects of changing the thermodynamic size while the kinetic dynamical region itself remains unchanged [20].

### *3.1 Phenomenological Bifurcations*

Figure 1 shows the top view projection of the 3-D stationary distribution of the Brusselator obtained from the maximum entropy approach according to the methodology described in [17]. This perspective makes it clear that as $k_2$ is decreased the distribution, which is initially unimodal, develops a crater of nearly zero probability and resembles the macroscopic limit cycle [21]. These same parameters in the macroscopic limit display non-oscillatory behavior for larger values of $k_2$ and sustained oscillations for smaller ones. Adopting Arnold's nomenclature for externally disturbed systems, the intrinsically stochastic Brusselator goes through a P-bifurcation as the set of kinetic parameters is varied.

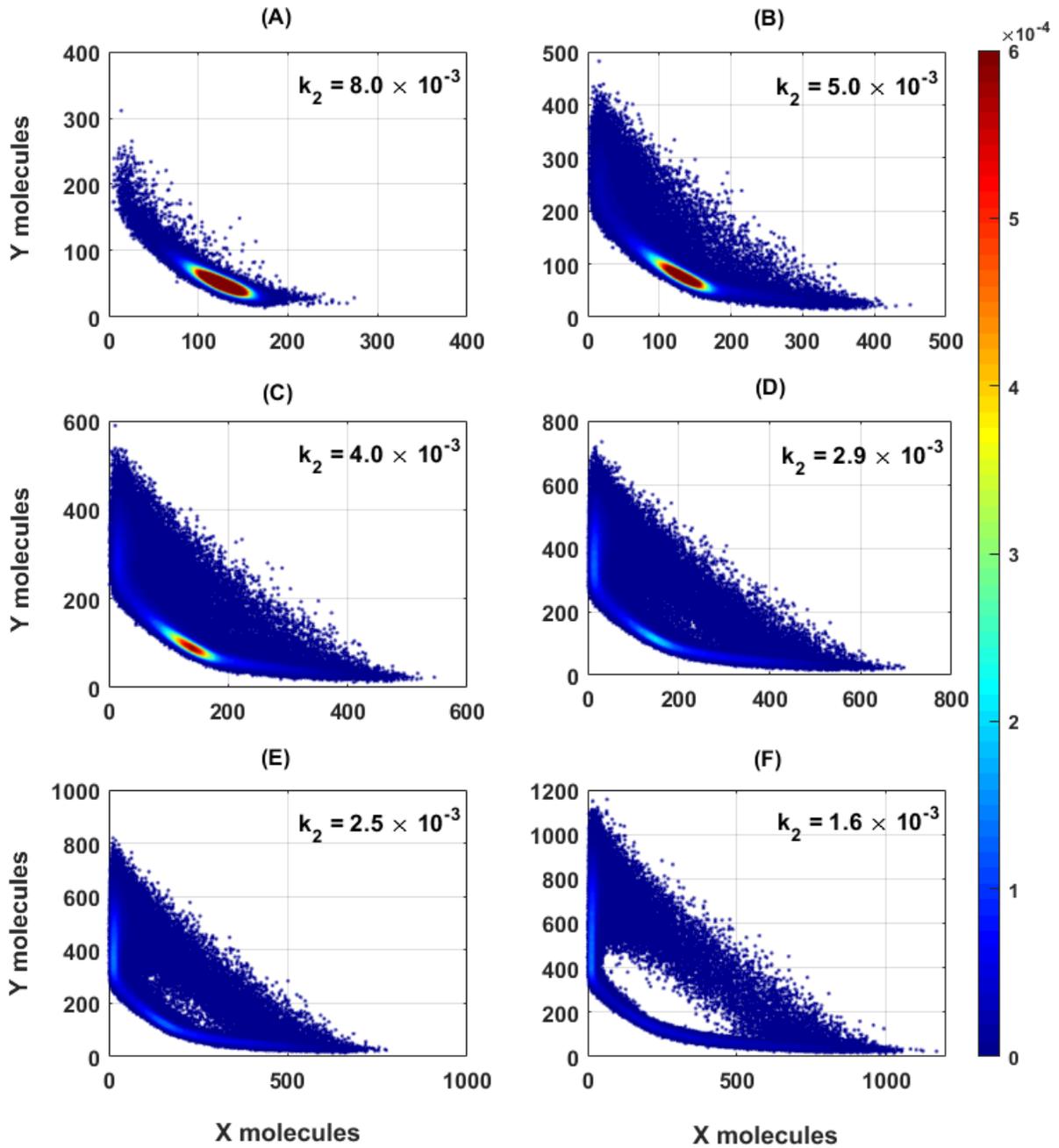

**Figure 1 | Brusselator Stationary Distribution versus Kinetic Parameter.** Top view projection of the 3-dimensional stationary distributions for the Brusselator model with different kinetic parameters $k_2$, while $k_1 = 625, k_3 = 50, k_4 = 5$ and $\Omega = 1.0$. Note that the same color map showing the probability is used for all sets.

More interestingly, however, Figure 2 shows that smaller system sizes can also cause a topological change in the distribution and ultimately destroy the crater-shape that characterizes the oscillatory region. This phenomenon is like the reverse Hopf bifurcation that Bashkirtseva et al. [9] observed in the van der Pol oscillator with multiplicative external noise. Nonetheless, in the current case the noise is internal and the distribution does not return to its original unimodal shape.

It is important to emphasize that for all systems shown in Figure 2, the set of kinetic parameters corresponds to the oscillatory region in the macroscopic limit. The mere reduction of the size of the systems undoes the crater formation, which sufficiently qualifies as a new type of phenomenological bifurcation.

*3.2 Hopf Dynamical Bifurcation*

The dynamics of the first four moments of the system with different kinetic constants are shown in Figure 3. These were obtained from kinetic Monte Carlo simulations using the software Hy3S based on Gillespie's stochastic simulations algorithm (SSA) with a million reaction trajectories [22-23]. The ensemble averages (probability moments) reach a steady state regardless of the parametric region. There is usually an initial overshoot followed by damped oscillations reaching a stationary distribution. However, as the systems transition from the non-oscillatory to the fully oscillatory region at the trajectory level ($k_2 = 2.9 \times 10^{-3}$ at the macroscopic limit), the moments themselves show stronger oscillations before reaching steady state. In the oscillatory region, the moments become more weakly damped and their relaxation time increases because the ensemble of trajectories becomes less easily coherent in terms of phases and amplitudes.

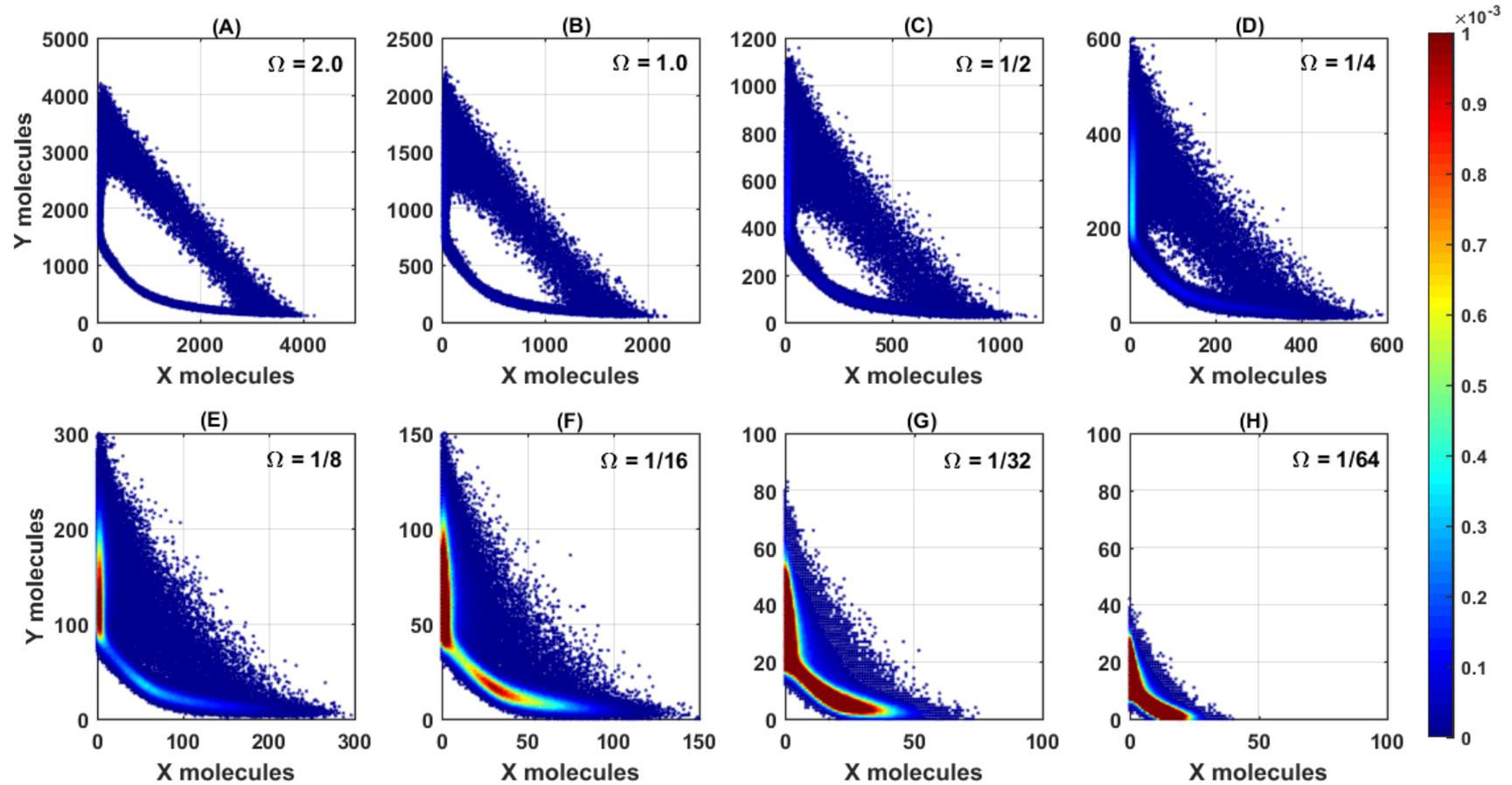

**Figure 2 | Brusselator Stationary Distribution versus Size Parameter.** Top view projection of the 3-dimensional stationary distributions for the Brusselator model with different size parameters $\Omega$, while $k_1 = 1250$, $k_2 = 4 \times 10^{-4}$, $k_3 = 50$ and $k_4 = 5$. Note that the same color map showing the probability is used for all sets.

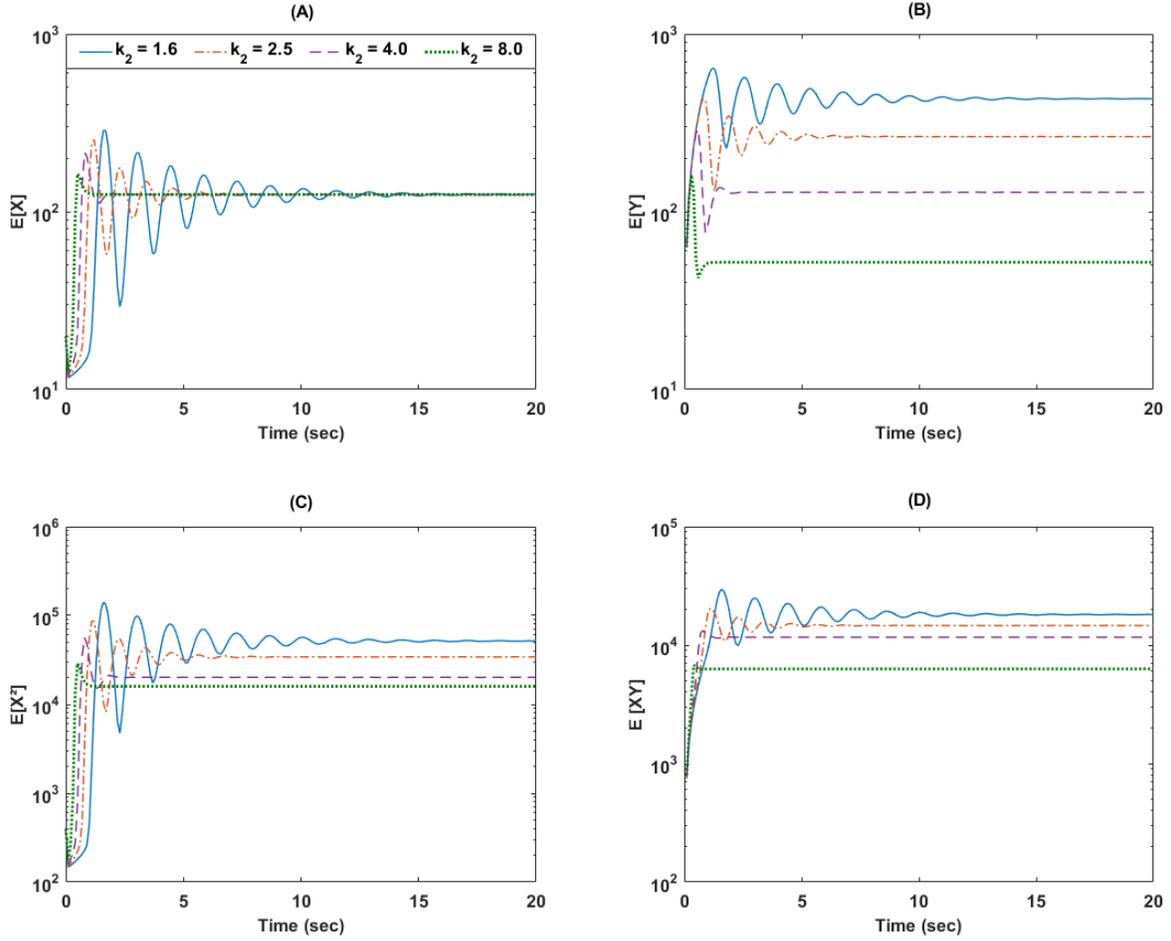

**Figure 3 | Kinetic Parameter Effect on the Brusselator Moment Dynamics.** Brusselator dynamics of the first 4 moments (E[X], E[Y], E[$X^2$], E[XY]) for different values of $k_2$, when $k_1 = 625$, $k_3 = 50$, $k_4 = 5$. In the legend of set (A), all $k_2$ parameters have been multiplied by $10^3$ for better visualization.

By plotting the real and imaginary components of the dominant eigenvalue of the stationary Jacobian matrix of the Brusselator (Figure 4), it is clear that the moment dynamics becomes more oscillatory as we decrease the kinetic parameter. The dominant eigenvalue crosses the imaginary axis and becomes unstable (Fig 4.B), somewhat similarly to what we observe in the deterministic Brusselator. The analytical solution for the dominant eigenvalue of the deterministic Brusselator according to Eqs. (5) through (7) is shown in Figure 4.A. In the macroscopic limit, as $k_2$ is decreased the stationary solution goes from stable node to unstable node. This change from stable to unstable is what mainly characterizes the deterministic Hopf bifurcation. In the stochastic

Brusselator, the solution goes from a stable node to an unstable node and then, if $k_2$ is decreased even further, an unstable focus.

Therefore, the main difference is that in the stochastic version the real part becomes unstable first and only then the imaginary axis is crossed. In the deterministic, the imaginary axis is crossed first and then the real part becomes unstable. However, in both cases the dynamical system presents a proper Hopf bifurcation. It is, nonetheless, incorrect to affirm that the stationary moments become unstable because as we observe from Figure 3 they reach a stationary state. However, the relaxation becomes much slower and oscillatory, which explains the overshoot and later damped oscillations in the moment dynamics.

These observations are only possible due to the ability to calculate moment eigenvalues through the maximum entropy closure scheme. To the best of our knowledge, such calculations for intrinsically stochastic systems with oscillatory behavior have never been done before. Evidently, one must be cautious in drawing parallels between deterministic and stochastic bifurcations because the deterministic refers to the stability of the steady state solution, which we can see as a single noise-free trajectory. The stochastic analysis, on the other hand, refers to the stability of the stationary moments, representing the ensemble of trajectories.

Nonetheless, results in Figure 4 demonstrate that by changing the kinetic parameters of the stochastic Brusselator, the system not only goes through a P-bifurcation but also through a proper dynamical bifurcation characterized by the change of sign in the real part of its dominant eigenvalue and the presence of a nonzero imaginary component.

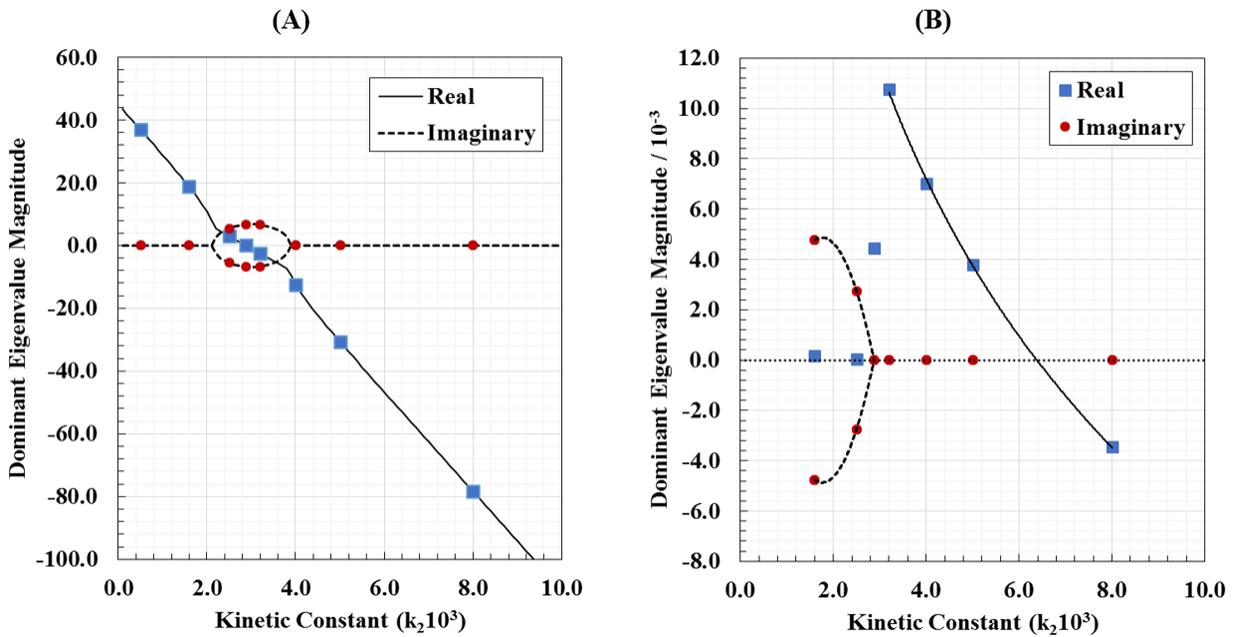

**Figure 4 | Dominant Eigenvalue versus Kinetic Parameter.** Both real (blue squares) and imaginary (red circles) components of the dominant eigenvalue are displayed. (A) Deterministic model in the macroscopic limit; Continuous and dashed lines represent the analytical solution, while the discrete points represent the corresponding systems that were chosen to be simulated stochastically in the microscopic limit. (B) Stochastic model using maximum entropy closure for the moment equations. The continuous curve represents the best fit based on the coefficient of determination ($R^2$) for the real part of the dominant eigenvalue ($y = -15.41 \ln(x) + 28.56$; $R^2 = 0.9996$). Dashed curve is only meant to guide the eye.

*3.3 Reverse Hopf Dynamical Bifurcation*

Similar to Figure 3, the dynamics of the first four moments of the system shown in Figure 5 starts out with strong oscillations and as the size parameter is decreased the moments reach steady state more rapidly and in a more stable fashion. Even though the system is in the oscillatory region based on the kinetic parameters chosen, the smaller system sizes make the moment dynamics behave similarly to the non-oscillatory region, such as the green dotted curves presented in Figure 3. This observation led us to conclude that the phenomenological bifurcation seen in Figure 2 could also be a proper dynamical bifurcation involving the size parameter. These results suggested that the Hopf bifurcation created by kinetic changes could be reversed and undone by simply decreasing the system size.

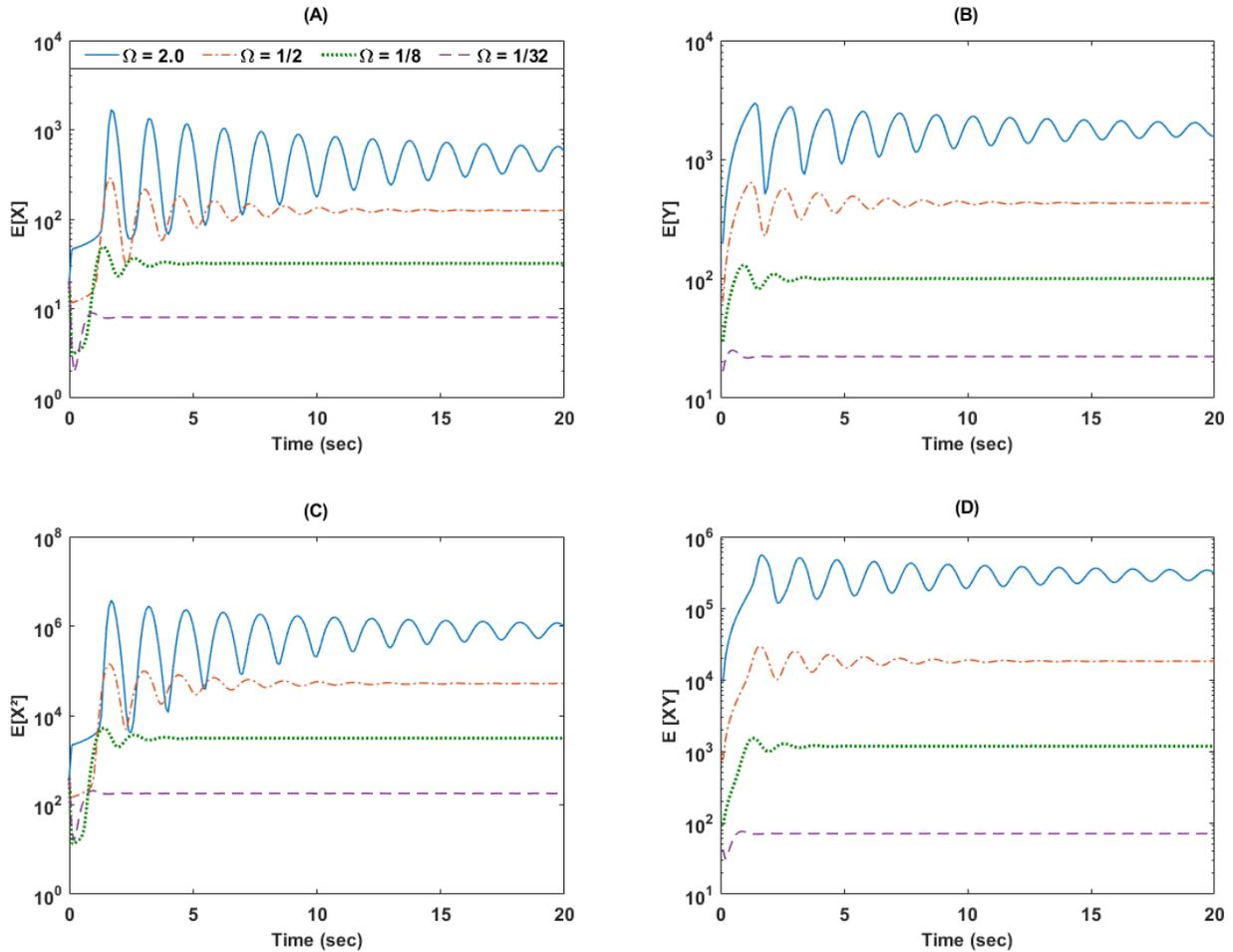

**Figure 5 | Size Parameter Effect on the Brusselator Moment Dynamics.** Brusselator dynamics of the first 4 moments ($E[X], E[Y], E[X^2], E[XY]$) for different values of Ω, when $k_1 = 1250$, $k_2 = 4 \times 10^{-4}, k_3 = 50$ and $k_4 = 5$.

With the maximum entropy closure scheme and the stability theory for stochastic systems derived thereof, we were able to confirm the phenomenon. Figure 6 shows that as the size of the system in the oscillatory parametric region decreases, the dominant eigenvalues of the moment equations become progressively less imaginary, until they are completely real. More interestingly, the real part, which is generally positive in the far oscillatory region, becomes negative, which indicates that the stationary solution grows increasingly more stable. This sufficiently characterizes a D-bifurcation, because a quantitative dynamical change is observed in the sign and domain space of the eigenvalues.

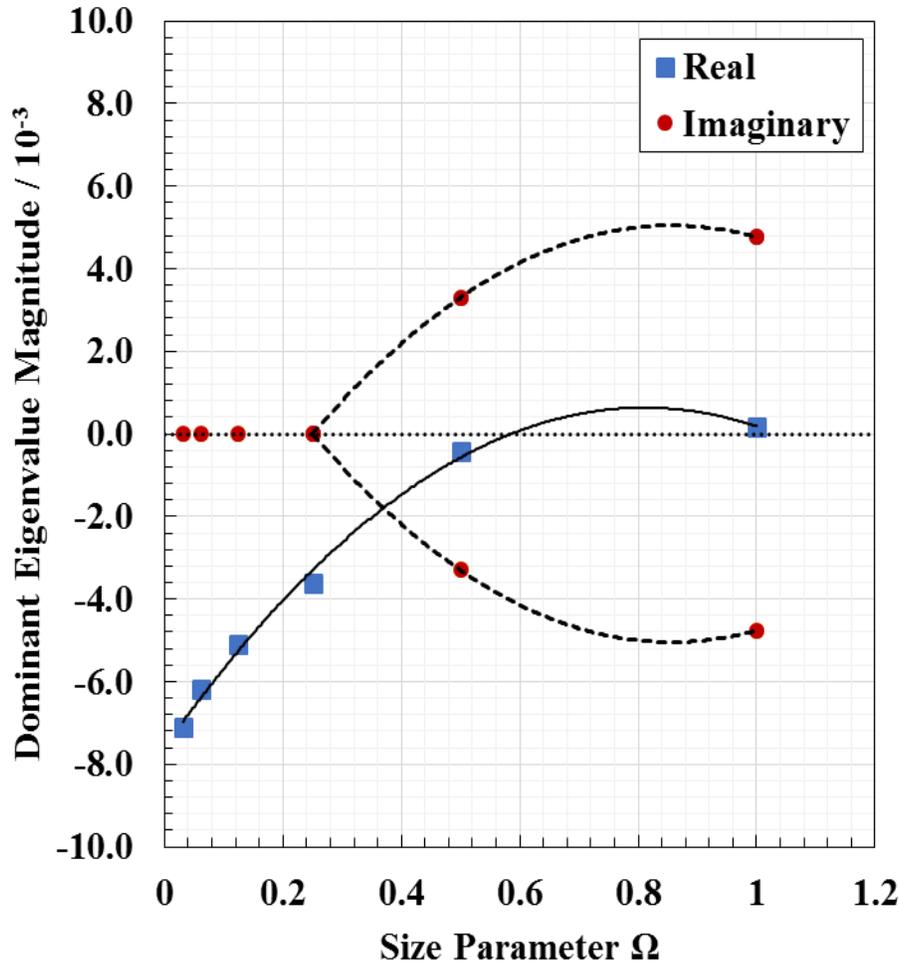

**Figure 6 | Dominant Eigenvalue versus Size Parameter $\Omega$.** Both real (blue squares) and imaginary (red circles) components of the dominant eigenvalue are displayed for $k_1 = 1250$, $k_2 = 4 \times 10^{-4}$, $k_3 = 50$ and $k_4 = 5$. The continuous curve represents the best fit based on the coefficient of determination ($R^2$) for the real part of the dominant eigenvalue ($y = -12.51x^2 + 20.277x - 7.583$; $R^2 = 0.9954$). Dashed curve is only meant to guide the eye.

At the trajectory level, the Hopf D-bifurcation with kinetic changes is hardly visualized by simply plotting the trajectory evolution in time (Figure 7.A-C). Regardless of the kinetic parameter, the amplitudes of the signals seem to be aperiodic. However, on the phase plane representation (Figure 7.D-F), we observe more clearly the crater development in the oscillatory region (smaller $k_2$), as some points in the phase space cease to be visited.

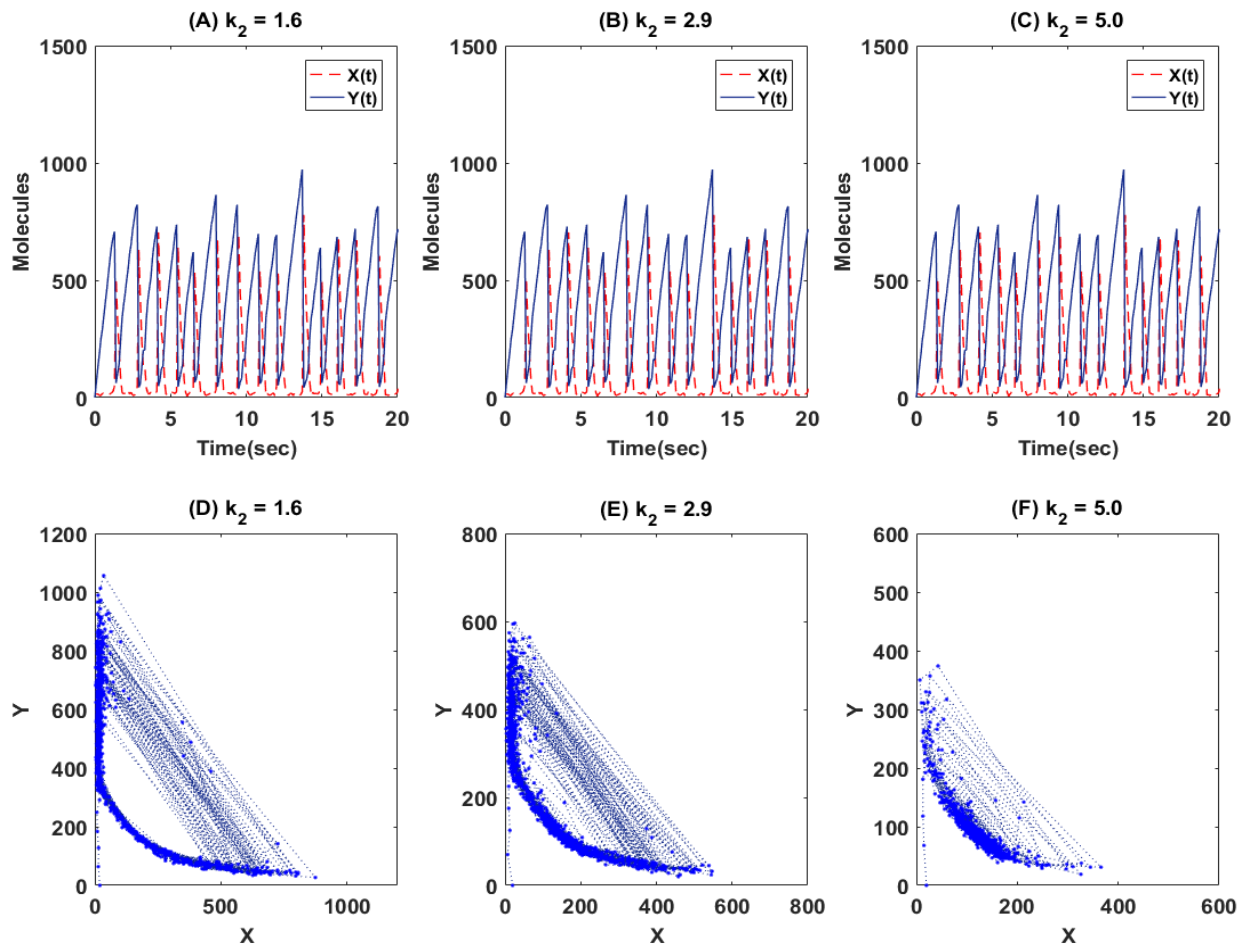

**Figure 7 | Sample SSA Trajectory with Different Kinetic Parameter $k_2$.** Red traced curve represents the X component, while the blue continuous curve displays the Y component for $\Omega = 1.0$. All $k_2$ parameters have been multiplied by $10^3$ for better visualization. (A)-(C) show the dynamical trajectory, while (D)-(F) show their respective phase planes.

The Discrete Fourier Transform (DFT) of the signals in Figure 7 are shown in Figure 8 and they allow us to determine whether there are main frequencies of oscillation that can be separated from thermal noise. We see that as the kinetic parameter is reduced, the system transitions to the oscillatory region and distinct peaks appear in the DFT signals. The further into the oscillatory region, the bigger the intensity of the peaks for both components X and Y.

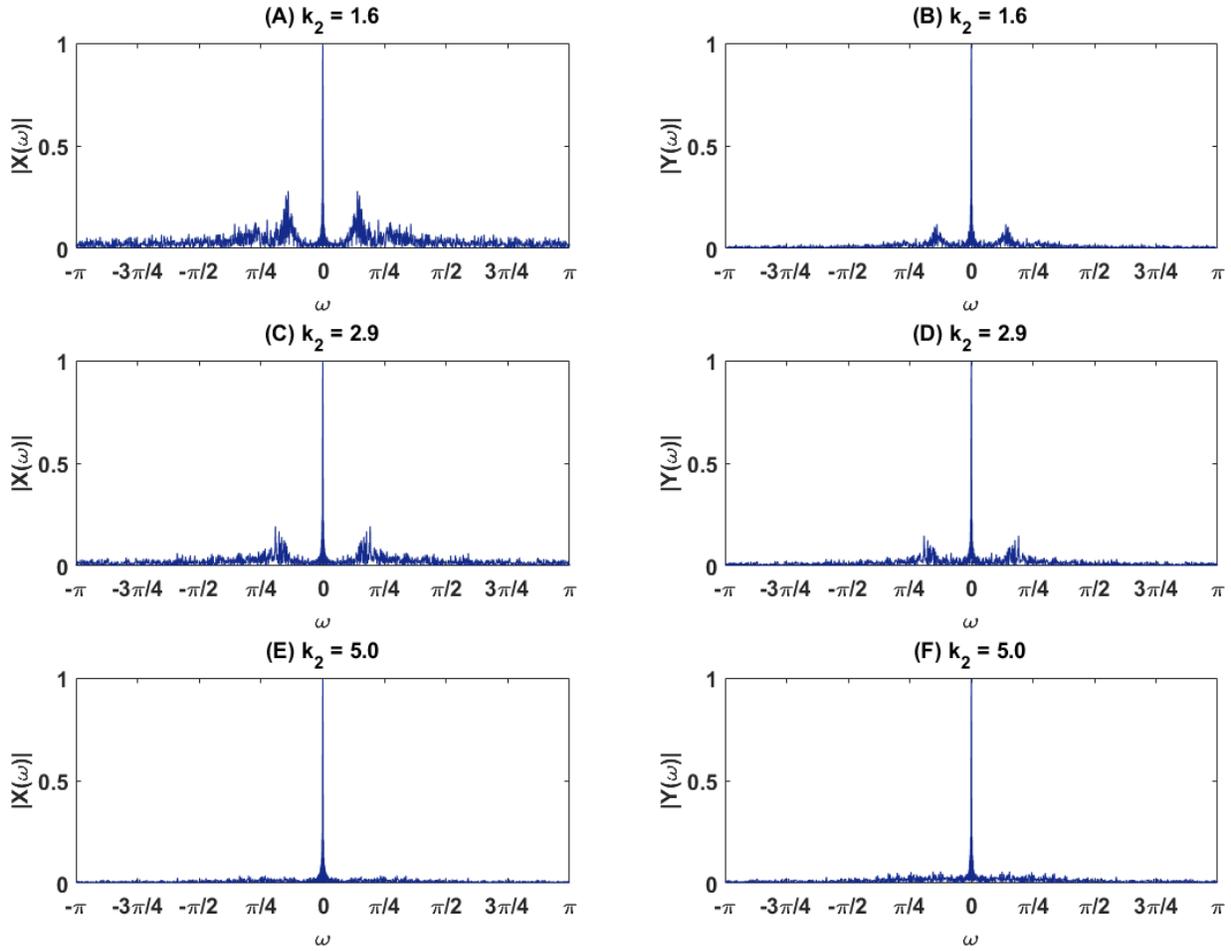

**Figure 8 | Discrete Fourier Transform of Sample Trajectory with Different Kinetic Parameter $k_2$.** All $k_2$ parameters have been multiplied by $10^3$ for better visualization. Left column (A, C, E) shows the DFT of the signal for component X, while the right column (B, D, F) is for component Y.

The reverse Hopf D-bifurcation with size changes is displayed at the trajectory level as oscillations being suppressed by the increasing thermal noise at smaller scales. The dispersion or variance in phases and amplitudes of each individual trajectory is so high that no limit cycle can be distinguished (Figure 9).

This suppression of oscillations by the intrinsic noise is further confirmed by the DFT signals shown in Figure 10, which are like the plots in the non-oscillatory region shown in Figure

8. The distinct peaks characterizing the main frequencies of oscillation disappear at small smaller system sizes for both X and Y components.

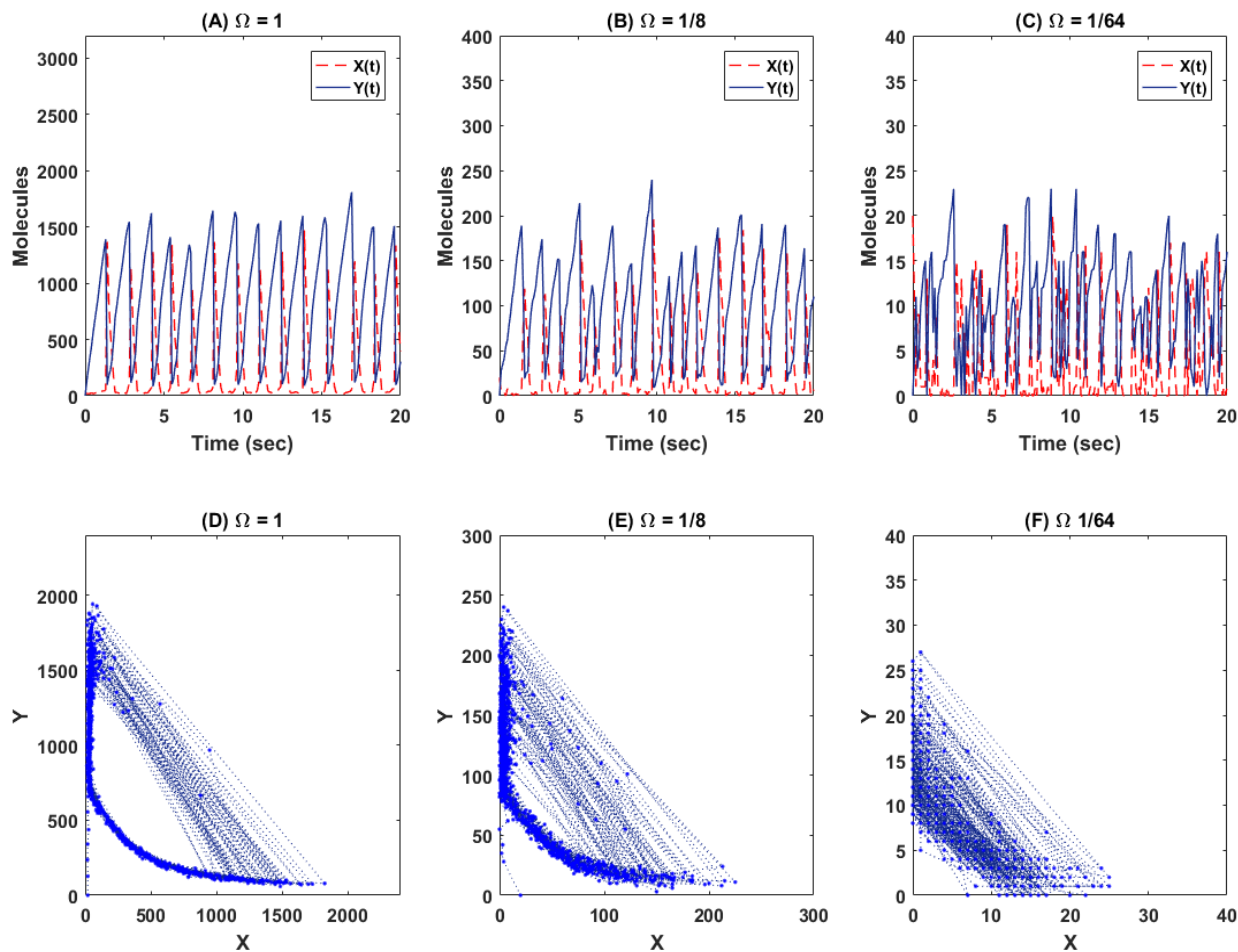

**Figure 9 | Sample SSA Trajectory with Different Size Parameter $\Omega$.** Red traced curve represents the X component, while the blue continuous curve displays the Y component for $k_1 = 1250$, $k_2 = 4 \times 10^{-4}$, $k_3 = 50$ and $k_4 = 5$. (A)-(C) show the dynamical trajectory, while (D)-(F) show their respective phase planes.

These results lead us to conclude that reducing system size is somehow equivalent to the large external parametric multiplicative noise experiments by Bashkirtseva et al. [9], in which the bifurcation shift was proportional to the square of the noise intensity. Here, the same is suggested by the best fitting curve to the functional dependency of the real component of the dominant eigenvalue with the size parameter, which is also quadratic. Both large external parametric multiplicative noise and large internal noise due to small size can reverse the Hopf bifurcation by destroying the crater development at the ensemble level and suppressing individual oscillations at the trajectory level.

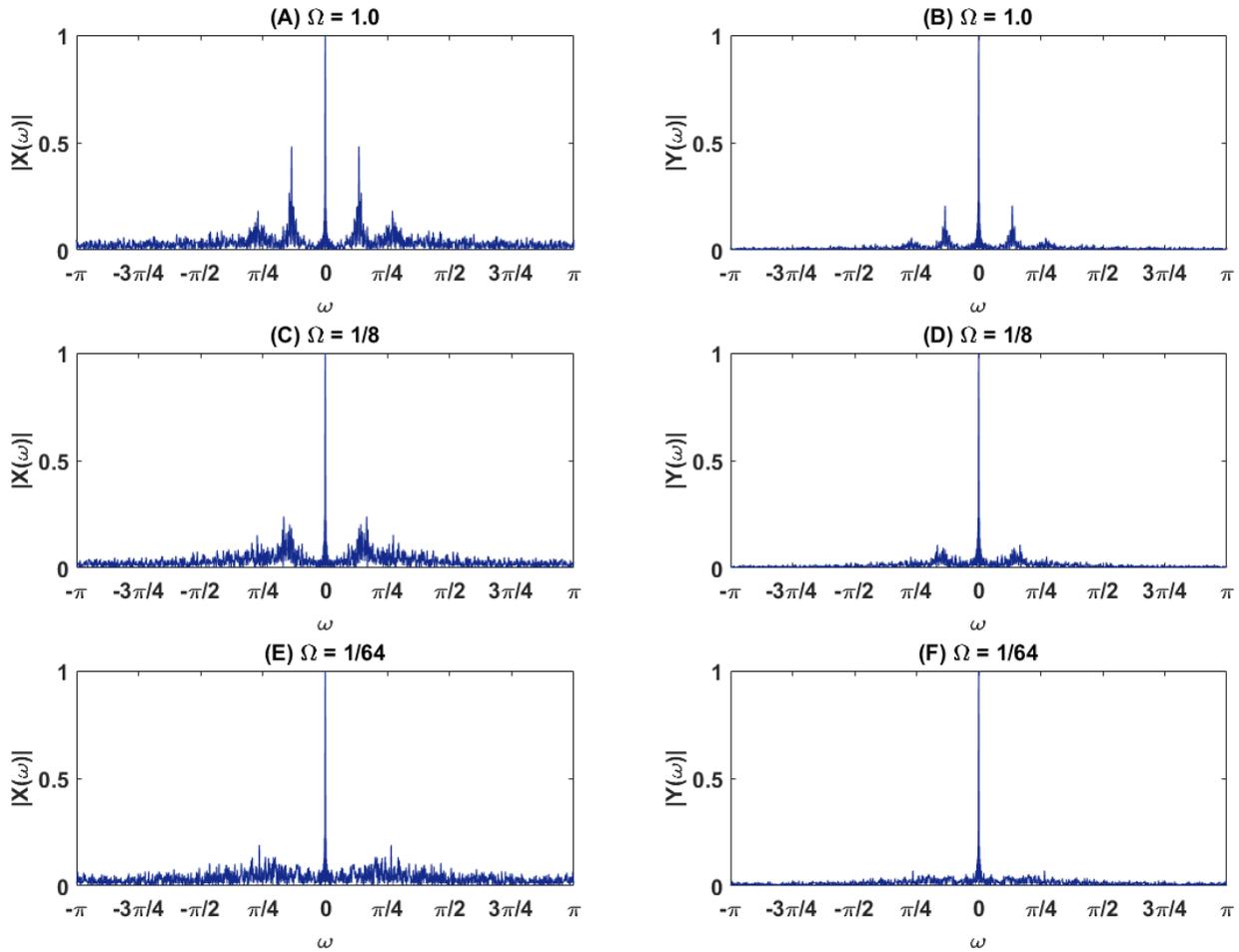

**Figure 10 | Discrete Fourier Transform of Sample Trajectory with Different Size Parameter Ω.** Kinetic parameters were $k_1 = 1250$, $k_2 = 4 \times 10^{-4}$, $k_3 = 50$ and $k_4 = 5$. Left column (A, C, E) shows the DFT of the signal for component X, while the right column (B, D, F) is for component Y.

## 4. Conclusions

In this paper we investigated the kinetic and size effects on the stochastic dynamics of the Brusselator system. We confirmed that as the kinetic parameters are changed from the non-oscillatory to the oscillatory region, the system develops a phenomenological bifurcation characterized by the appearance of a crater-like shape inside the stationary distribution. Kinetic Monte Carlo simulations were employed to track the dynamics of the moments, which showed overshooting and stronger oscillations before reaching steady state as the parametric region became oscillatory. These stochastic simulations results were then explained by the application of the maximum entropy closure scheme for moment equations and the stability analysis proposed. The stationary solution transitions from a stable node to a stable spiral and this behavior was captured by the dominant eigenvalue of the moment equation Jacobian. To the extent of our knowledge, this is the first time that linear stability analysis has been applied to internally stochastic systems with oscillatory dynamics. The eigenvalue analysis allowed us, therefore, to reach the bigger conclusion that the change in kinetic parameters generates a proper dynamical bifurcation.

Regarding size effects, our numerical experiments have shown that even if the system is in the oscillatory parametric region, the distribution displays significant topological changes as the system becomes smaller. Monte Carlo simulations showed that in such cases the crater shape that is usually described in the literature as the main feature of oscillatory systems can also be destroyed through another phenomenological bifurcation. The stability analysis stablished that this bifurcation is also dynamical. As the system size decreases, the stationary solution becomes increasingly stable at the ensemble level. This phenomenon was captured by the moment dynamics and the dominant eigenvalue becoming more negative and real. The stationary solution transitions from a stable spiral to a stable node. This reversing of the Hopf bifurcation even while the system is in the oscillatory parametric region has never been discussed in the literature and it is unique to the stochastic model. We believe that the results here presented exemplify the applicability of the theoretical framework for stability analysis of stochastic systems based on the maximum entropy closure scheme.

Furthermore, our results point forward to the possibility of development of model predictive control strategies for biochemical oscillators in artificial circuits (synthetic biology) and biomedical applications. For instance, time-delayed transcription-translation negative feedback produces sustained oscillations in gene transcription at the cellular level within each tissue involved in mammalian circadian rhythms (e.g., regulation of glucose metabolism in the liver) [24]. It has been recently proposed that the damping rate of population-level bioluminescence recordings of cultured circadian reporter cells can serve as an accurate measure of stochastic noise [25]. Therefore, the eigenvalue analysis shown here for the ensemble dynamics of the moments can also be used as a measure of stochastic noise in the system. In principle, eigenvalues could quantify the effectiveness of small-molecule therapeutics in maintaining high-amplitude circadian rhythms or in manipulating phase resetting for metabolic health, as these modulators change the damping rate [24, 26].

**Acknowledgments:** This work was supported by a grant from the National Science Foundation (CBET-1412283) and a grant from the National Institutes of Health (GM111358). This work utilized the high-performance computational resources of the Extreme Science and Engineering Discovery Environment (XSEDE), which is supported by National Science Foundation grant number ACI-1053575. Support from the University of Minnesota Digital Technology Center and from the Minnesota Supercomputing Institute is gratefully acknowledged. This paper was written in part while YNK was a Visiting Scholar at the Isaac Newton Institute of Mathematical Sciences at the University of Cambridge. Support from the University of Minnesota Biotechnology Institute and CAPES – Coordenação de Aperfeiçoamento de Pessoal de Nível Superior – Brazil is also acknowledged.

**Author Contributions:** Conceptualization, P.H.C. and Y.N.K.; Methodology, P.H.C.; Software, P.H.C.; Validation, P.H.C.; Formal Analysis, P.H.C.; Investigation, P.H.C.; Resources, Y.N.K.; Data Curation, P.H.C.; Writing – Original Draft Preparation, P.H.C.; Writing – Review & Editing, P.H.C. and Y.N.K.; Visualization, P.H.C.; Supervision, Y.N.K.; Project Administration, Y.N.K.; Funding Acquisition, Y.N.K.

**Conflicts of Interest:** The authors declare no conflict of interest.